\documentclass{osa-article}
\usepackage{amsmath}
\usepackage{amsfonts}
\usepackage{mathrsfs}
\journal{oe}

\begin{document}

\title{Differential real-time single-pixel imaging with Fourier domain regularization - applications to VIS-IR imaging and polarization imaging}

\author{Anna Pastuszczak,\authormark{1,*} Rafa{\l} Stojek,\authormark{1,2}, Piotr Wr{\'o}bel,\authormark{1} and Rafa{\l} Koty{\'n}ski\authormark{1}}

\address{\authormark{1}University of Warsaw, Faculty of Physics, Pasteura 5, 02-093 Warsaw, Poland\\
\authormark{2}Vigo System, Pozna{\'n}ska 129/133, 05-850 O{\.z}arów Mazowiecki, Poland}

\email{\authormark{*}apastuszczak@igf.fuw.edu.pl} 




\begin{abstract}
The speed and quality of single-pixel imaging (SPI) are fundamentally limited by image modulation frequency and by the levels of optical noise and compression noise. 
In an approach to come close to these limits, we introduce a SPI technique, which is inherently differential, and comprises a novel way of measuring the zeroth spatial frequency of images and makes use of varied thresholding of sampling patterns.
With the proposed sampling, the entropy of the detection signal is increased in comparison to standard SPI protocols.  Image reconstruction is obtained with a single matrix-vector product so the cost of the reconstruction method scales proportionally with the number of measured samples. A differential operator is included in the reconstruction and following the method is based on finding the generalized inversion of the modified measurement matrix with regularization in the Fourier domain. 
We demonstrate $256 \times 256$ SPI at up to 17 Hz at visible and near-infrared wavelength ranges using two polarization or spectral channels. A low bit-resolution data acquisition device with alternating-current-coupling can be used in the measurement indicating that the proposed method combines improved noise robustness with a differential removal of the direct current component of the signal.
\end{abstract}

\section{Introduction}
Indirect image measurement techniques called single-pixel imaging (SPI), since their introduction over a decade ago~\cite{Duarte_2008,Cand_s_2006}, have led to a considerable amount of novel ideas about image measurement at various wavelength ranges, spectral imaging, imaging through scattering media, 3D imaging etc.~\cite{Gibson2020,Edgar_2018}.

The speed of single-pixel imaging is limited by the frequency of spatial modulation and by the time needed for digital image reconstruction. Most real-time reconstruction approaches rely on a single-step image reconstruction using a fast transform, e.g. the Fourier (FFT), Walsh-Hadamard (FWHT), or discrete cosine (DCT) transforms, or on the evaluation of a  matrix-vector product\cite{Czajkowski_2018,Stantchev2020}. There is also growing interest in using neural networks for image reconstruction or for removing artifacts caused by compression\cite{Higham2018,Rizvi2020}.

Modulation speed obtained with modern digital micromirror devices (DMD), such as the one used by us, is on the order of tens of kilohertz. This is not a lot, as many thousand exposures are needed per single image measurement.
Far higher frame-rates are achievable with structured illumination with LEDs arrays, although such set-ups for ghost imaging and SPI have been demonstrated only at low resolutions~\cite{Salvador-Balaguer2018,Xu2018,Wang_2020,Zhao_2019}. Modulation with rotary elements with fixed patterns is a high-speed cost-efficient alternative to using dynamic modulators in THz\cite{Guerboukha_2018,Chen2019}.

Block compressive imaging also effectively increases the sampling frequency per pixel with use of multiple detectors or a focal plane-array\cite{Ke_2012,Mahalanobis_2014,Wu_2019}.

Differential photodetection techniques allow to eliminate some of the optical noise, and to effectively increase the SNR of the detection signal. Differential imaging and normalized imaging have been used in computational ghost imaging prior to SPI\cite{Ferri_2010,Sun_2012}. 
In SPI, a reference detector may be used either to measure the unmodulated reference signal and to normalize the  detection signal in the presence of source intensity fluctuations, or to measure in parallel a differently modulated signal. The latter possibility is often based on the fact that DMD micromirrors may take exactly one of the two positions (let us call them "0" or "1"). Intensities of two reflected beams are proportional to the result of image sampling with complementary (binary negated) patterns\cite{Yu2014,Radwell_2014}. Their difference provides a signal with improved SNR and with removed bias. This is possible when the collimated incident beam comes from a perpendicular direction onto the DMD, for instance in a telescopic set-up\cite{Yu2014}. Otherwise the set-up must be carefully  calibrated to retain a similar level of signal from both detectors, which becomes challenging for 3D moving and rotating objects. Still, this kind of balanced photodetection has been used by many groups\cite{Soldevila2016,Czajkowski_2018,2019OExpr..27.4562D}.

A more straightforward way to obtain differential detection is to subtract the measurements from two subsequent measurements\cite{Welsh2013}. This is a way not only to remove  a slowly changing bias from the signal  but also to encode both positive and negative values of real-valued sampling functions\cite{Zhao_2015,Lan_2016,Pastuszczak_2016,Yu_2016}. A more elaborated encoding allows to represent complex-valued sampling functions with three\cite{Zhang2017} instead of four\cite{Zhang_2015} real-valued non-negative patterns. This approach has its origins in interferometry and also allows to remove a constant-bias at the same time. Encoding a complex pattern with a larger number of non-negative intensity signals has been also proposed\cite{De_Tommasi_2016}. A simplex encoding may also be used to represent an arbitrary bunch of real-valued sampling functions using non-negative patterns whose number is increased by one. Once again the constant bias of the measurement is removed, and by varying the simplex dimension one may select an optimal trade-off between optical noise and compression noise, depending on the experimental conditions\cite{Czajkowski_2019}. The same method may be easily combined with complementary detection to improve the SNR even further\cite{Czajkowski_2019}. Encoding  real or complex valued functions onto non-negative intensity patterns usually does not produce binary patterns yet, and a further binarization step is necessary. For instance error diffusion is commonly used with success thanks to the difference in DMD resolution and the actual imaging resolution.

In this paper we introduce  a differential method for measuring the DC component of images (the zeroth spatial frequency) applicable to various SPI architectures with binary modulation.
In a classical approach, the DC component is measured using a single pattern filled with all pixels in the state "1", possibly followed by another pattern with all  pixels in the state "0"\cite{Yu2014,Radwell_2014}.
The proposed measurement method makes use of a small number of binary patterns with approximately (but not exactly) half of DMD pixels in the "1" and "0" states to encode the DC component. 
 This allows to limit the required operating range of the detector and to improve the noise robustness of the measurement. Unlike in \cite{De_Tommasi_2016,Czajkowski_2019},  the sampling is readily binary. At the same time it retains the advantageous of other differential methods such as the immunity of the measurement to some constant or slowly varying bias of the detection signal. 
On top of this, we propose a complete fast differential SPI framework with binary sampling and a single step image reconstruction. The set of sampling patterns includes several patterns required to capture the DC component of the image, and other patterns taken from some standard basis but binarized with variable thresholding. The purpose of varying the binarization level is to make the best use of the available detector measurement range. Image reconstruction is based on the Fourier domain regularized inversion (FDRI)\cite{Czajkowski2018} with a minor modification that includes the differential treatment of the measurement signal. We will use the short name D-FDRI for the modified method. In a simple comparison made under experimental conditions, the proposed reconstruction method gives a similar image quality as optimization with a compressive sensing method\cite{NESTA_2011}.
Finally, we validate D-FDRI experimentally. Concurrent imaging in the infrared and visible spectral bands, likewise polarisation imaging are fields where SPI is an interesting alternative to classical optical approaches\cite{Edgar_2015,Gehm_2015,Gibson_2017,Li_2018,Yao_2018,Duran_2012,Soldevila_2013,Welsh_2015,Seow_2020}. Making use of rather standard single pixel camera set-ups we demonstrate the capability of conducting a real time continuous measurement and reconstruction at a resolution of $256\times256$ in two independent channels, using moderate computational resources. 

\section{Differential SPI with Fourier-domain regularized inversion}
Compressive measurement of an image $\mathbf{x}$ (with pixel values arranged in a vector) in the presence of additive signal noise $\mathbf{n_s}$ and detector noise $\mathbf{n_d}$ may be  expressed with a matrix-vector multiplication
\begin{equation}
\mathbf{y}=\textbf{M}\cdot (\mathbf{x+n_s})+\mathbf{n_d} \label{eq.measurement},
\end{equation}
where the rows of the measurement matrix $\mathbf{M}$ contain the patterns displayed on the DMD during the measurement, and $\cdot$ is the dot product. In~Eq.~(\ref{eq.measurement}) noise has been decomposed into a signal independent part $\mathbf{n_d}$ primarily attributed to the detector dark current and the signal dependent part $\mathbf{n_s}$ from sources such as background illumination\cite{Sun_2018}. The measurement is compressive when the dimension of $\mathbf{y}$ which will be denoted as $k$ is smaller than the number of pixels $n$ in the image $\mathbf{x}$. In our case $n=256^2$ and the compression ratio $CR=k/n\geq2\%$. 

Consider applying discrete difference operator of the order p  on $\mathbf{y}$,
\begin{equation}
\mathbf{y'}=\mathbf{D}^p_k\cdot \mathbf{y}= (\mathbf{D}^p_k\cdot \mathbf{M})\cdot (\mathbf{x+n_s}) +\mathbf{D}^p_k\cdot\mathbf{n_d}.\label{eq.diffmeasurement}
\end{equation}
Matrices $\mathbf{D}^p_k$ have the size of $(k-p)\times k$ and are defined as $[\mathbf{D}_k^1]_{i,j}=\delta_{i,j-1}-\delta_{i,j}$ (discrete gradient operator) and $[\mathbf{D}_k^2]_{i,j}=\delta_{i,j-1}-0.5\delta_{i,j-2}-0.5\delta_{i,j}$ (discrete Laplace operator), where $\delta$ is the Kronecker delta symbol. 

Applying the gradient or Laplace operators on $\mathbf{y}$ makes $\mathbf{y'}$ invariant to the level of constant bias present in the detection signal represented in our model by the mean value of $\mathbf{\bar{n}_d}$. In the case of Laplace operator, a bias changing linearly with time is also removed. In practice, using $\mathbf{y'}$ for image reconstruction allows to operate the data acquisition device (DAQ) with alternating-current (AC) coupling at a reduced voltage range with the bit-resolution better adapted to cover the signal range.

The generalized inverse of the matrix product $\mathbf{D}^p_k\cdot \mathbf{M}$ multiplied by $\mathbf{y'}$ gives an approximation for image $\mathbf{x}$,
\begin{equation}
\tilde{\mathbf{x}} = {(\mathbf{D}^p_k\cdot \mathbf{M})^{g} }\cdot(\mathbf{D}^p_k\cdot \mathbf{M})\cdot \mathbf{x}=\mathbf{P_g}\cdot \mathbf{y},
\end{equation}
where $\mathbf{P_g}=(\mathbf{D}^p_k\cdot \mathbf{M})^g \cdot\mathbf{D}^p_k$ and $~^g$ denotes the generalized matrix inverse, which is not unique.  The most commonly used generalized matrix inverse operator is the Moore-Penrose (pseudo)inverse. Pseudoinverse is known to minimize the mean square error (MSE) in linear regression and is sometimes used in single-pixel imaging. However, in a compressive measurement, it may be more advantageous to use a different kind of matrix generalized inverse. For instance based on Fourier-domain regularization (FDRI)\cite{Czajkowski_2018}, the solution to Eq.~(\ref{eq.measurement}) may be expressed as $\mathbf{x_0}=\mathbf{P}\cdot \mathbf{y}$ with
\begin{equation}
\mathbf{P}=\mathbf{F^*}\cdot\mathbf{\hat\Gamma}\cdot\mathbf{F}\cdot(\mathbf{M}\cdot\mathbf{F^*}\cdot\mathbf{\hat\Gamma}\cdot\mathbf{F})^{+},
\end{equation}
where $^+$ denotes the pseudoinverse, and $\mathbf{F}$ is the 2D Fourier transform.   $\mathbf{\hat\Gamma}$ is a diagonal matrix
\begin{equation}
[\mathbf{\hat\Gamma}]_{i,j}=\frac{\delta_{i,j}}{\sqrt{(1-\mu)^2(\sin^2(\omega_x)+\sin^2(\omega_y))+\mu^2\frac{\omega_x^2+\omega_y^2}{2\pi^2}+\epsilon }}\label{eq.Gamma},
\end{equation}
where $\mu$ and $\epsilon$ are used to tune the properties of the regularization\cite{Czajkowski_2018} and $\omega_{x,y}$ are the spatial frequencies. 
Taking into account the Laplace or gradient operators from Eq.~(\ref{eq.diffmeasurement}) which effectively modify the measurement matrix, the following matrix may be used to reconstruct the image in a differential measurement,
\begin{equation}
\mathbf{P_g}=\mathbf{F^*}\cdot\mathbf{\hat\Gamma}\cdot\mathbf{F}\cdot(\mathbf{D}^p_k\cdot\mathbf{M}\cdot\mathbf{F^*}\cdot\mathbf{\hat\Gamma}\cdot\mathbf{F})^{+}\cdot\mathbf{D}^p_k.\label{eq.P_g}
\end{equation}
Calculation of matrix $\mathbf{P_g}$ is computationally costly, and in practice only possible at a small compression ratio $CR=k/n$. 
On the other hand, the measurement matrix $\mathbf{M}$ as well as $\mathbf{P_g}$ can be precalculated before the measurement. We use the singular-value decomposition (SVD) to calculate the pseudoinverse in Eq.~(\ref{eq.P_g}), and the FFT algorithm to calculate the 2D Fourier transforms. Then the image reconstruction requires us to evaluate a single matrix-vector product $\tilde{\mathbf{x}}=\mathbf{P_g}\cdot \mathbf{y}$. Additionally, any negative values in the reconstructed image are replaced with zeros. The cost of matrix-vector multiplication is  $O(k\cdot n)$. For $n=256\times256$-pixel images sampled at a compression ratio of $CR=2\%$, a real-time ($17$~Hz) reconstruction does not require using GPU-accelerated computation, even when two independent channels are present. In fact, we have developed a program in Python based on integer and floating-point single-precision arithmetic capable of receiving, interpreting and averaging data from the DAQ and reconstructing images in two bands in real time on an Intel i5-4590 CPU. Also at a higher compression ratio, e.g. when $CR=6\%$ with a lower level of compression noise, the operating speed is limited by the DMD frame-rate and not by digital processing.

In comparison to reconstructions with inverse transforms, for which there exist fast algorithms such as FFT, DCT or FWHT, the proposed D-FDRI method is slower and more memory demanding.
Its advantages are that it works well with arbitrary sampling matrices $\mathbf{M}$, also with nonorthogonal and binary patterns.
The computational cost of FFT is $O(n\cdot \log2 n)$. On the other hand, that of the matrix-vector multiplication is $O(k\cdot n)$ or $O(CR\cdot n^2)$. Therefore in practice the method is applicable at small compression ratios $CR$. The reconstruction properties, including noise robustness, may be further optimized by the choice of $\mu$ and $\epsilon$, and the reconstructed image is invariant to the presence of a constant or linearly changing in time bias in the detection signal.

\subsection{Differential measurement of the DC-component with binary sampling functions} 
The usual method of measuring the zeroth spatial frequency (the DC-component) of the image is to display a pattern consisting of all pixels equal to "1" on the DMD, possibly followed by a pattern consisting of all pixels equal to "0". We  have replaced a direct measurement with a measurement distributed over several sampling functions. 
 We wanted them to be binary, with approximately but not exactly half of the pixels in states "0" and "1". On top of this, information about the DC component of the image should be preserved after applying the finite difference operator on the measurement with these patterns.

\begin{figure}[h!]
	\centering\includegraphics[width=13cm]{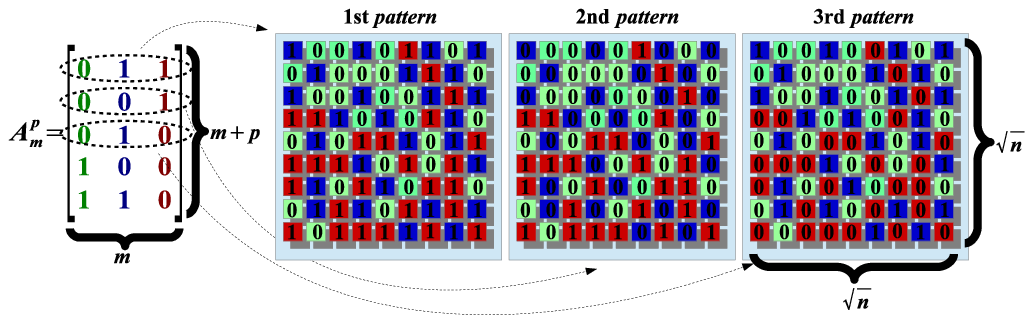}
	\caption{This figure explains how the binary sampling patterns are created from matrix $\mathbf{A}^p_m$. We divide the DMD into $m$ classes of pixels. Every class  contains approximately $n/m$ randomly chosen DMD pixels (shown with distinct colors).  Rows of $\mathbf{A}^p_m$ encode $m+p$ binary patterns that enable to recover the DC component of the image in a differential measurement.
	Each element of the row governs the state of one class of pixels within a single pattern. \label{fig.matrixmapping}}
\end{figure}

To this end, we group the DMD pixels into an odd number of classes $m$. The class number is randomly assigned to every pixel. To encode the measurement of the zeroth spatial frequency of the image we define $m+p$ binary sampling patterns. Each of these patterns consists of either $(m+1)/2$ or $(m-1)/2$ pixel classes in state "1". We note that such patterns contain approximately half of the pixels in states "0" and "1". To find a possible mapping between the  pixel classes and the patterns we define a  small binary matrix $\mathbf{A}_m^p$ of size $(m+p)\times m$. Each row of the matrix defines a single sampling pattern, and each column represents one class of pixels.
We performed a numerical search for such binary matrices with the condition that every row of the matrix contains $(m\pm 1)/2$ equal values (zeros or ones) and that $rank(\mathbf{D}^p_{m+p}\cdot\mathbf{A}^p_m)=m$. This assures that a row consisting of ones is linearly dependent with the rows of $\mathbf{D}^p_{m+p}\cdot\mathbf{A}^p_m$, implying that information about the DC component is contained in the measurement also after applying the Laplace or gradient operator $\mathbf{D}^p_{m+p}$. Among many possible solutions, we have chosen such matrices $\mathbf{A}^p_m$ that minimize the dispersion of coefficients needed to reconstruct the DC component from the measurement with $\mathbf{D}^p_{m+p}\cdot\mathbf{A}_m^p$. As an example, the following matrices satisfy these conditions for $m=3$ and $m=7$ and for the order of the finite difference operator $p=1$ and $p=2$,

\begin{equation}
\mathbf{A}_3^1 = 
\begin{pmatrix}
0 & 1 & 1 \\
1 & 1 & 0 \\
1 & 0 & 0 \\
0 & 1 & 0
\end{pmatrix},
\mathbf{A}_3^2 = 
\begin{pmatrix}
0 & 1 & 1 \\
0 & 0 & 1 \\
0 & 1 & 0 \\
1 & 0 & 0 \\
1 & 1 & 0
\end{pmatrix},
\end{equation}
\begin{equation}
\mathbf{A}_{7}^1 = 
\begin{pmatrix}
0& 1& 0& 0& 1& 1& 1\\
1& 1& 0& 0& 0& 0& 1\\
0& 0& 1& 1& 1& 0& 0\\
1& 0& 1& 1& 0& 0& 1\\
0& 1& 1& 0& 0& 1& 0\\
0& 1& 0& 1& 1& 0& 1\\
1& 1& 1& 0& 0& 0& 1\\
0& 0& 1& 0& 1& 1& 1
\end{pmatrix},
\mathbf{A}_{7}^2 = 
\begin{pmatrix}
0& 0& 0& 0& 1& 1& 1\\
1& 0& 0& 1& 1& 0& 0\\
1& 1& 0& 0& 1& 0& 0\\
1& 1& 1& 1& 0& 0& 0\\
1& 0& 0& 1& 0& 1& 0\\
0& 1& 0& 1& 1& 0& 1\\
0& 0& 1& 0& 1& 1& 1\\
0& 0& 1& 0& 0& 1& 1\\
0& 1& 1& 1& 0& 0& 0
\end{pmatrix}.
\end{equation}After finding $\mathbf{A}^p_m$ and mapping its columns to groups of DMD pixels, $m+p$ additional binary sampling patterns are created. Figure~\ref{fig.matrixmapping} graphically illustrates how the binary sampling patterns are constructed with the help of matrix $\mathbf{A}^p_m$. In Fig.~\ref{fig.matrixmapping}, three colors indicate the three pixel classes. Every DMD pixel is randomly assigned to exactly one of the classes, and pixels within each class take the same value equal to the corresponding element of matrix $\mathbf{A}^p_m$.

\begin{figure}[h!]
	\centering\includegraphics[width=13cm]{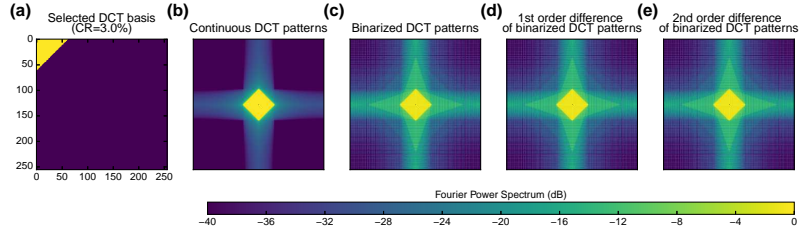}
	\caption{(a)~Selected subset of low-frequency DCT patterns (shown in yellow), (b)~Power spectral density (PSD) of the real-valued DCT sampling functions, (c)~PSD of the binarized DCT sampling functions, (d)(e)~PSD of the binarized DCT sampling functions to which the first~(d) or second~(e) order difference operator was applied.\label{fig.spectrum}}
\end{figure}

\begin{figure}[h!]
	\centering\includegraphics[width=13cm]{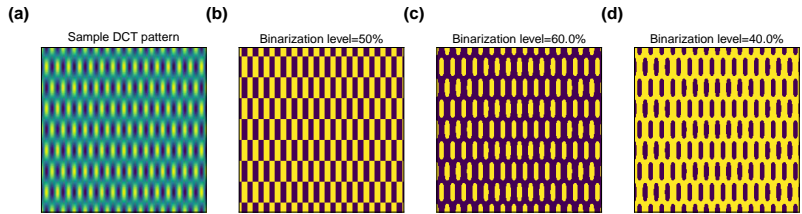}
	\caption{Binarization of a real-valued basis pattern at various levels. (a)~sample DCT real-valued pattern; (b)(c)(d)~the same pattern binarized at the levels of $50\% (b), 60\%$ (c), and $40\%$ (d). The binarization level determines the percentage of DMD pixels in the state "0" for the binarized pattern.\label{fig.binarization}}
\end{figure}
\begin{figure}[h!]
	\centering\includegraphics[width=15cm]{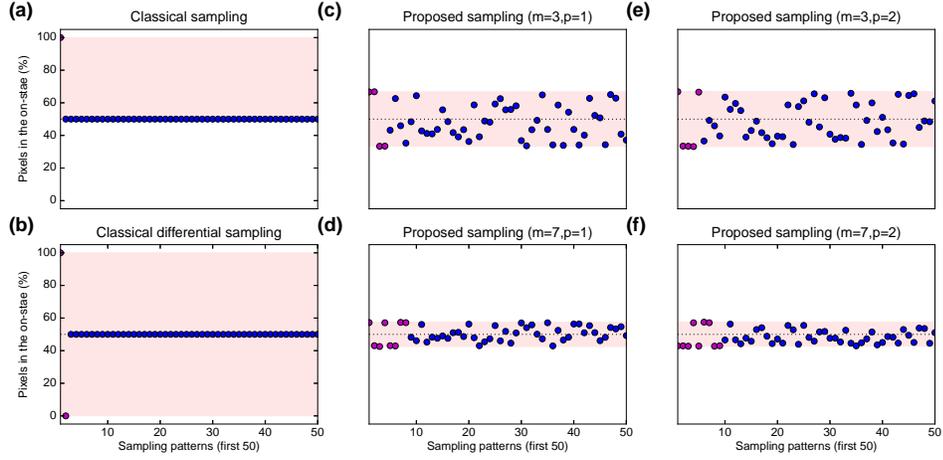}
	\caption{The fraction of DMD pixels in state "1" in the first 50 sampling patterns. Patterns used to measure the DC-component come first (and are plotted in magenta), other e.g. binarized DCT patterns follow (and are plotted in blue).  (a)~classical sampling with the DC image component measured with a single pattern with all pixels in state "1", and with half pixels in both states for the rest of patterns, (b)~classical differential sampling with the DC image component measured with two patterns with all pixels in the states "1" and "0". (c)(d)(e)(f)~The proposed differential sampling protocols for selected values of $p$ and $m$. The DC of the image is measured using $p+m$ binary patterns, and DCT patterns are binarized at a random level in the range $[(m-1)/2m, (m+1)/2m]$.\label{fig.sampling}}
\end{figure}

The patterns used in the measurement of the image DC-component are included as the first patterns in the measurement matrix $\mathbf{M}$. Other patterns are chosen from some selected basis.  We use a subset of binarized DCT basis corresponding to the lowest spatial frequencies. These components are selected in a way shown in Fig.~\ref{fig.spectrum}(a).  The binarization level is selected randomly in the range $[(m-1)/2m, (m+1)/2m]$. By the binarization level we understand the percentage of pixels in state "0" after binarization. A sample DCT pattern binarized at various levels is shown in Fig.~\ref{fig.binarization}, while Fig.~\ref{fig.sampling}  illustrates the effect of randomizing the binarization level on the number of pixels in state "1" for the proposed sampling in comparison with a classical sampling protocol. SPI measurement of a uniform image would produce a detection signal $\mathbf{y}$ with the same distribution as that shown in Fig.~\ref{fig.sampling}. An image dominated with low-frequency contents is likely to give a similar signal as well, and the binarization range gives a rough indication of the operating range of the detector.

  Fig.~\ref{fig.spectrum}(c) shows that binarization of the real-valued sampling patterns enhances the high spatial frequency contents of their power spectral density (PSD). On the other hand, Fig.~\ref{fig.spectrum}(c)(d) indicate that the PSD is little affected by the finite difference operators. The effective measurement matrix $\mathbf{D}^p_k\cdot\mathbf{M}$ consists of linear combinations of patterns in $\mathbf{M}$, and in terms of PSD, the differential sampling is nearly equivalent to the original one consisting of binarized DCT patterns.

\subsection{Tuning the method} 
The properties of the D-FDRI framework presented in this section depend on the choice of several parameters, namely $p,m,\epsilon$ and $\mu$. 

 The value of $m$ determines the number of additional samples that probe the DC component of an image and the thresholding range used in the binarization of the sampling functions. We used $m=7$ most of the time, which implies that the binarization level of sampling functions is selected randomly in the range $[43\%, 57\%]$, although values of $m$ between $3$ and $11$ also work well. As a rule of thumb, the thresholding level should be close to $50\%$, but varying it, increases the entropy of the measured signal, and improves noise robustness of the method.

The order of the differential operator $p\in\lbrace 1,2\rbrace$ is not as important as we expected. We did not observe any practical advantage of using $p=2$ over $p=1$ in the presence of  noise  under experimental conditions.
 When $p=2$ the calculation of pseudoinverse in Eq.~(\ref{eq.P_g}) usually resulted in obtaining some very small eigenvalues in the SVD. Still, in practice we reached a similar imaging quality with both operators. As a remark, Eq.~(\ref{eq.P_g}) should be evaluated using double-precision floating point arithmetic, even if $\mathbf{P_g}$ is later stored with a lower precision for the reconstruction stage. 
 
 \begin{figure}[h!]
	\centering\includegraphics[width=10cm]{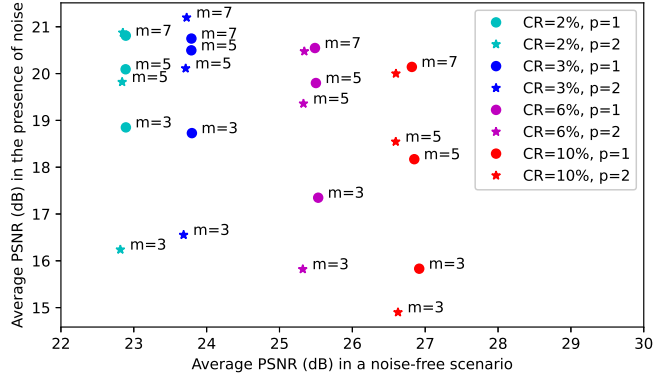}
	\caption{Average PSNR obtained with the proposed differential sampling technique without noise vs. average PSNR obtained in the presence of additive white noise with $\sigma=0.004$, where $\sigma$ is the standard deviation of the noise divided by the range of the measured SPI signal. The plot compares results obtained for compression ratios of $2\%, 3\%, 6\%, 10\%$, for parameter $m=3,5,7$ and $p=1,2$, when $\epsilon=10^{-7}$ and $\epsilon=0.5$. The simulation includes the presence of white additive Gaussian noise with standard deviation $\sigma R$, where $R$ denotes the range of the measured SPI signal. PSNR values are averaged over results obtained for 5120 images from the Caltech256 image database.\label{fig.comparison}}
\end{figure}

In Fig.~\ref{fig.comparison} we demonstrate how the parameters $m$ and $p$ influence the performance of the D-FDRI method without noise and in the presence of measurement noise. To measure the quality of the reconstructed images, we use the peak signal-to-noise ratio (PSNR) metric defined as:
\begin{equation}
PSNR({\bf \tilde x, x}) = 10 \log_{10} \frac{\max({\bf x})^2}{\frac{1}{n} \sum_{i=1}^{n} (x_i - \tilde x_i)^2},
\label{eq.psnr_def}
\end{equation}
where $\max({\bf x})$ is the maximal pixel brightness of the original image.
PSNR is affected both by the measurement noise and compression noise. Without measurement noise PSNR primarily depends on the compression CR. However at a certain level of experimental noise,  PSNR can no longer be increased by increasing CR. This situation is illustrated in Fig.~\ref{fig.comparison}, which shows the drop in PSNR due to the measurement noise for various values of $p$ and $m$. Noise robustness improves with the increase of $m+p$. In fact, PSNR explicitly depends on the mean value of the image, which is found from the measurement with $m+p$ patterns. The more patterns are used to determine the DC component, the lower is its measurement uncertainty.

The significance of parameters $\epsilon$ and $\mu$ in Eq.~(\ref{eq.Gamma}) has been discussed in Ref.~\cite{Czajkowski_2018}. The two parameters require fine-tuning. While $\mu\approx0.5$ works reasonably well in most situations~\cite{Czajkowski_2018}, the value of $\epsilon$ may significantly affect noise robustness of the method. It is reasonable to take $\epsilon<<1$ and then to test by trial and error the value of $\epsilon$ and respective magnitude of the elements in matrix $\mathbf{P_g}$. In the case $|\mathbf{P_g}|$ contains elements with a large magnitude, image reconstruction with the formula $\tilde{\mathbf{x}}=\mathbf{P_g}\cdot \mathbf{y}$ would enhance the measurement noise.

\subsection{Complementary sampling}
SPI optical set-ups with complementary sampling make use of the two beams reflected from the DMD. A single measurement gives two measurement vectors, $\bf{y}_1=\bf{M}\cdot\bf{x}$ and $\bf{y}_2=(1-\bf{M})\cdot\bf{x}$ (where we have neglected the noise terms, and $\bf{1}$ stands for a $k\times n$ matrix filled with ones). The usual approach is to take the difference $\bf{y}=(\bf{y_1}-\bf{y_2})/2$ as the result of a differential measurement. This allows for increasing the SNR and removing an equal bias signal from the two detectors. 

Since in D-FDRI the differential operator is included in the definition of $\bf{P}_g$ in Eq.~(\ref{eq.P_g}), the reconstructed image does not change when some constant value $c$ is added to $\bf{y}$
\begin{equation}
    \bf{P}_g\cdot\bf{y}=\bf{P}_g\cdot(\bf{y}+c).
\end{equation}
This makes it possible to acquire data from the detectors with AC coupling. On the other hand, it allows to construct two independent data channels with the two measurements $\bf{y}_1$ and $-\bf{y}_2$. The same reconstruction matrix $\bf{P}_g$ may be used in the two channels. The final formula for image reconstruction in the two channels is therefore
\begin{equation}
    \tilde{\mathbf{x}_l}=\theta\left(\mathbf{P_g}\cdot \mathbf{y}_l\cdot (-1)^{l-1}\right),
\end{equation}
where $l=1,2$ is the channel number, and $\theta(x)=x\text{ if }x>0,\text{ and }0\text{ otherwise}$.
In the present work, images reconstructed in the two channels are represented with different colors.

\section{Imaging quality, noise robustness, and signal entropy}

Entropy of the signals measured by the SPI system is calculated as
\begin{equation}
H = - \sum_{i=1}^{2^b} P(v_i) \log_2 P(v_i),
\label{eq.entropy_def}
\end{equation}
where $P(v_i)$ is the probability of an outcome of a single measurement $y_j$ ($j=1,2,...,k$) taking value $v_i$. The total number of possible outcomes of a single measurement equals $2^b$, where $b$ is the number of bits used in the analogue-to-digital (A/D) conversion of the signal.

In the information theory, entropy is a common measure of the average information carried by the possible outcomes of a stochastic variable. The less probable a specific outcome is, the more information is conveyed in its occurrence. While entropy is limited by the number of possible states of the random event $H \leq b$, its maximal value is attained in case of events with uniform probability distribution $P(v_i)$. It may be intuitively rephrased that the outcome of an experiment is most uncertain, and therefore most informative, when all possible results are equally probable. Entropy is also an essential measure for data compression, as it represents the absolute limit of how well data may be compressed without loss.

\begin{figure}[h!]
	\centering\includegraphics[width=10cm]{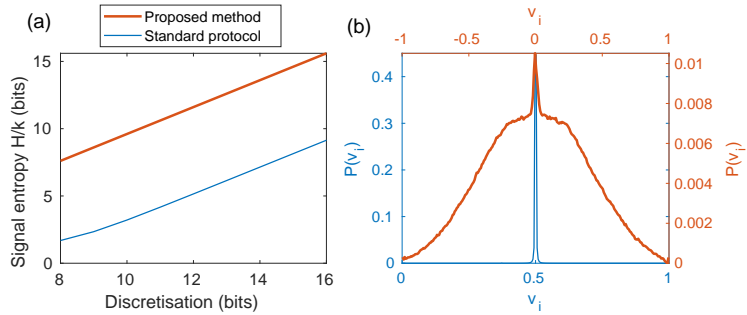}
	\caption{(a)~Average entropy per measurement pattern $H/k$ of the measured SPI signals as a function of the signal bit-resolution $b$. (b)~Experimental probability distributions of the occurrence of specific voltage values $v_i$ in the measured SPI signals for $b=8$. Results obtained from simulated measurements with over 180 thousand images.}
	\label{fig.entropy}
\end{figure}

In Fig.~\ref{fig.entropy}(a) we compare our proposed differential SPI sampling protocol with the standard non-differential one in terms of the average entropy per measurement $H/k$. The respective probability distributions $P(v_i)$ for both sampling methods are presented in Fig.~\ref{fig.entropy}(b).
We estimate the probability distributions $P(v_i)$ experimentally, based on the histograms of simulated measurements calculated for over 180 thousand images. The images are taken from the Caltech 256 Image Dataset~\cite{CALTECH256}, which contains 30 thousand images organized into 256 distinct categories. Additional data augmentation (rescaling, random cropping, flipping, rotation) was used to further increase the amount of data. For each of the images, we calculate the SPI signal $\bf y$ obtained by sampling the image with patterns stored in matrix $\bf M$, according to Eq.~(\ref{eq.measurement}) excluding the noise. Then, each signal $\bf y$ is normalized. For the proposed sampling, we assume that the mean value of the measurements $\bf \bar y=0$ (i.e. only the AC component of the signal is measured) and the signal fits in range ${-}1 \leq y_j \leq 1$ for $j=1,2,..,k$. For the classical sampling, all measurements are non-negative and, after normalization, they fit into the range $0 \leq y_j \leq 1$, where the maximal value 1 is practically obtained only for the measurement with the pattern which has all pixels in the "1" state. Finally, the signals are discretised using between 8 and 16 bits as $b$, which corresponds to the range of the bit-resolution of the DAQ. The estimates of $P(v_i)$ are calculated for each value of $b$ separately as the average probability of measuring value $v_i$ over all of the sampling patterns and the whole image dataset.

While the standard SPI protocol results in the concentration of the measured values $y_j$ around the center of the DAQ range, the proposed differential method yields much broader spread of the measured signal. This broadening is achieved thanks to the following properties of the proposed method: (1) the elimination of the constant bias from the measurements by introducing the differential measurement protocol, (2) decomposition of the DC measurement into several patterns with transparency close to (but not equal) $50 \%$, and (3) the application of randomized thresholding to the DCT functions. The resulting entropy of the proposed sampling is close to its upper limit and takes an average value $H/k \approx 0.96\ b$, while for the standard sampling protocol $H/k \approx 0.4 b$ on average.

For the purpose of numerical evaluation of the proposed method, in the present and the following simulations we have taken $m=7, p=2, \mu=0.5$ and  $\epsilon=10^{-8}$. The size of the scene is $n=256^2$, and the number of DCT sampling functions is equal to $k=1974+m+p$, which corresponds to the compression ratio $CR\approx3\%$. DCT basis corresponding to the lowest spatial frequencies are selected,  and they are  thresholded randomly to contain between $43\%$ and $57\%$ pixels in the "1" and "0" states. As a reference, we take the non-differential FDRI~\cite{Czajkowski_2018} with the same basis functions discretized at $50\%$, and with $\mu=0.5$. This way, the comparison is focused on the significance of including the differential operator  and on the varied thresholding in D-FDRI.

\begin{figure}[h!]
	\centering\includegraphics[width=13cm]{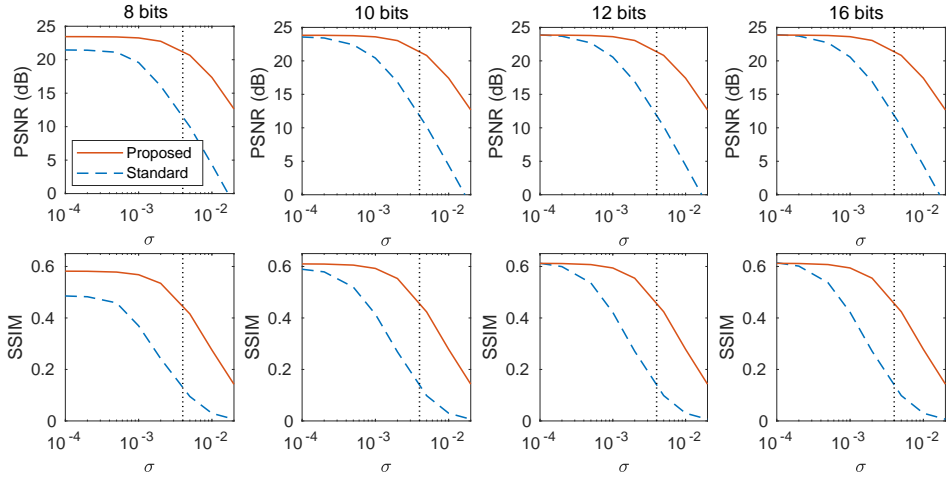}
	\caption{Sensitivity of the proposed sampling method (red) to the experimental noise $\mathbf{n}_d$. Direct DCT sampling (blue) for comparison. Average PSNR and SSIM values obtained from simulated SPI measurements and reconstructions with 5120 images. The simulation includes the presence of white additive Gaussian noise with $\sigma$ denoting the  standard deviation of the noise divided by the range of the measured SPI signal. This noise model corresponds to the inherent statistical noise of the SPI camera and it is independent of the A/D discretization noise. Respective subplots are obtained for $8$-bit, $10$-bit, $12$-bit, and $16$-bit  discretization.
	Vertical dotted lines mark the level of noise present in our experimental SPI set-up.}
	\label{fig.noise}
\end{figure}

\begin{figure}[h!]
	\centering\includegraphics[width=9cm]{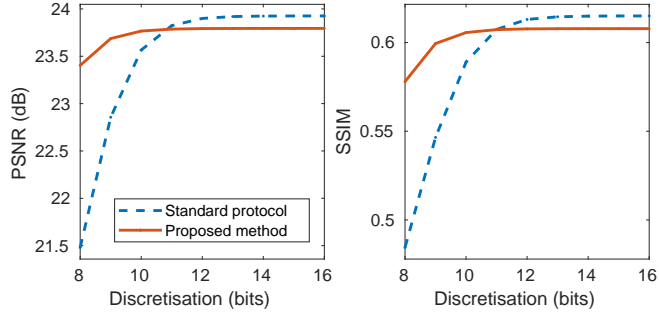}
	\caption{Sensitivity of the proposed sampling method (red) to the A/D discretization. Standard non-differential SPI sampling with the same set of DCT patterns (blue) for comparison. Average PSNR~(a) and SSIM~(b) values obtained from simulated SPI measurements and reconstructions with 5120 images. Before the image reconstruction, the SPI signals are converted into signals with b significant bits, i.e. only $2^b$ distinct values are measured by the detector. We use b between 8 and 16, which corresponds to the range of the bit-resolution of our experimental equipment. We assume no presence of other types of experimental noise.}
	\label{fig.discretisation}
\end{figure}
In Figs.~\ref{fig.noise}~and~\ref{fig.discretisation} we analyze the robustness of the proposed method to noise coming from two mutually independent sources: the statistical detector noise $\bf n_d$, modeled as additive white Gaussian noise, and the discretization noise applied to the signal at the stage of A/D conversion. 
For this reason, we have performed simulated measurements and reconstructions with a set of 5120 images selected randomly from the Caltech 256 database.
For evaluation of the quality of the reconstructed images, we use two metrics, namely: the PSNR defined in  Eq.~(\ref{eq.psnr_def})  and the mean similarity index (SSIM). 
 
For calculating SSIM, we use the built-in Matlab function \textit{ssim} with default parameters, which implements the mean SSIM metric from \cite{Wang2004}.

In contrast to the standard SPI protocol, D-FDRI is significantly less sensitive to both types of noise present in the SPI cameras. Especially its robustness to the discretization noise allows to lower the cost of the SPI set-up by taking advantage of a cheaper, lower-resolution DAQ with almost no impact on the quality of the recorded images.
We have found that in the case of using an 8-bit DAQ, the average decrease of PSNR due to the discretization  is $\sim 0.6$~dB for the proposed method, as compared to almost 2.5~dB for the standard non-differential sampling. 

\section{Experimental polarization imaging and VIS-IR imaging}
\begin{figure}[h!]
	\centering\includegraphics[width=13cm]{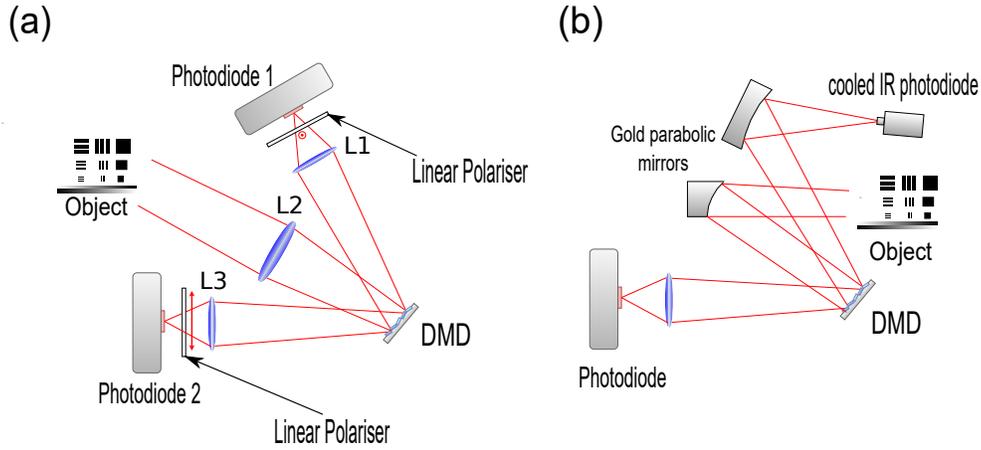}
	\caption{Schematics of the SPI configurations with two channels. (a)~Polarization imaging with complementary channels measuring orthogonal linear polarizations. (b)~Concurrent measurement of visible and near-infrared bands. }
	\label{fig.schem}
\end{figure}

In this section we validate the proposed D-FDRI SPI framework experimentally. 
The schematics of our optical SPI set-ups for polarization imaging and imaging in the visible and  near infrared wavelength bands are shown in Fig.~\ref{fig.schem}(a) and (b), respectively. The modulator splits the incoming optical beam into two complementarily modulated beams that are measured independently with two bucket detectors. In the case of polarization imaging, linear orthogonally oriented polarizers are put in front of the two photodiodes, and after image  reconstruction, the polarization information is indicated with pseudocoloring. In the presented situation, the scene is illuminated with unpolarized light which passes through a linear polarizer and a moving birefringent object. In the case of VIS-IR imaging, two different detectors are used for measuring the intensities of the beams reflected from the DMD.
 A similar set-up could allow for a complementary differential measurement, however here the two channels are used to measure independently two polarization states or two wavelength bands, respectively,  while the sampling and reconstruction method assure the differential properties of the measurement. 

\begin{figure}[h!]
	\centering\includegraphics[width=13cm]{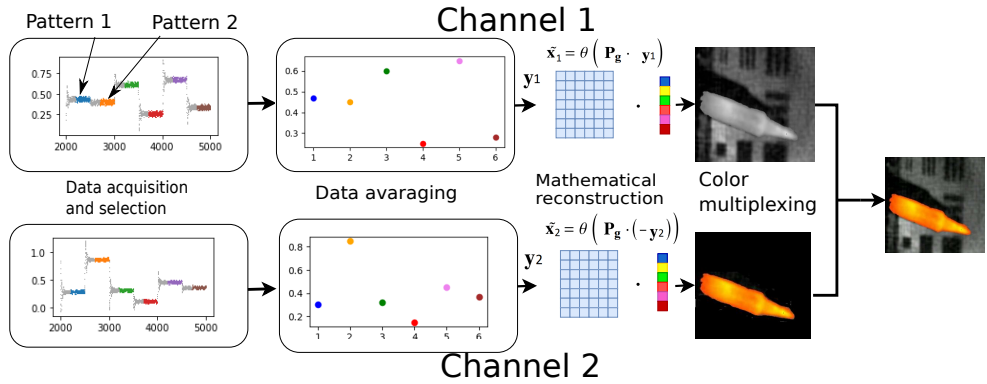}
	\caption{Flow-chart explaining the processing of acquired data. Data is acquired continuously at a sampling of $256$ns. The data gathered during the exposure of a single sampling pattern  (marked in color) are selected and averaged to obtain a single measurement $y_i$. Averaging is performed independently in each of the two channels. Then, monochromatic images from the two channels are reconstructed independently and integrated by color multiplexing to produce a single pseudocolor frame.}
	\label{fig.flowchart}
\end{figure}

\begin{figure}[h!]
	\centering\includegraphics[width=13cm]{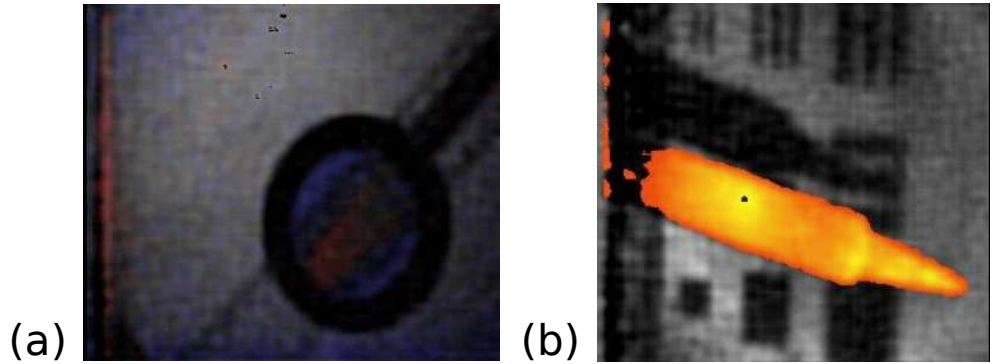}
	\caption{Sample pseudocolor frames obtained with the real-time two-channel SPI. (a)~Polarization imaging of a birefringent object (ruler) placed in front of a linear polarizer, (b)~VIS-IR imaging of a hot object (soldering iron tip at 400$^{\circ}$C). Respective visualisations are \textbf{Visualization1}- showing polarization imaging of a rotated linear polarizer, \textbf{Visualization2}-with a moving birefringent object, \textbf{Visualization3}-with the moving soldering iron tip.}
	\label{fig.lutownica}
\end{figure}

\begin{figure}[h!]
	\centering\includegraphics[width=13cm]{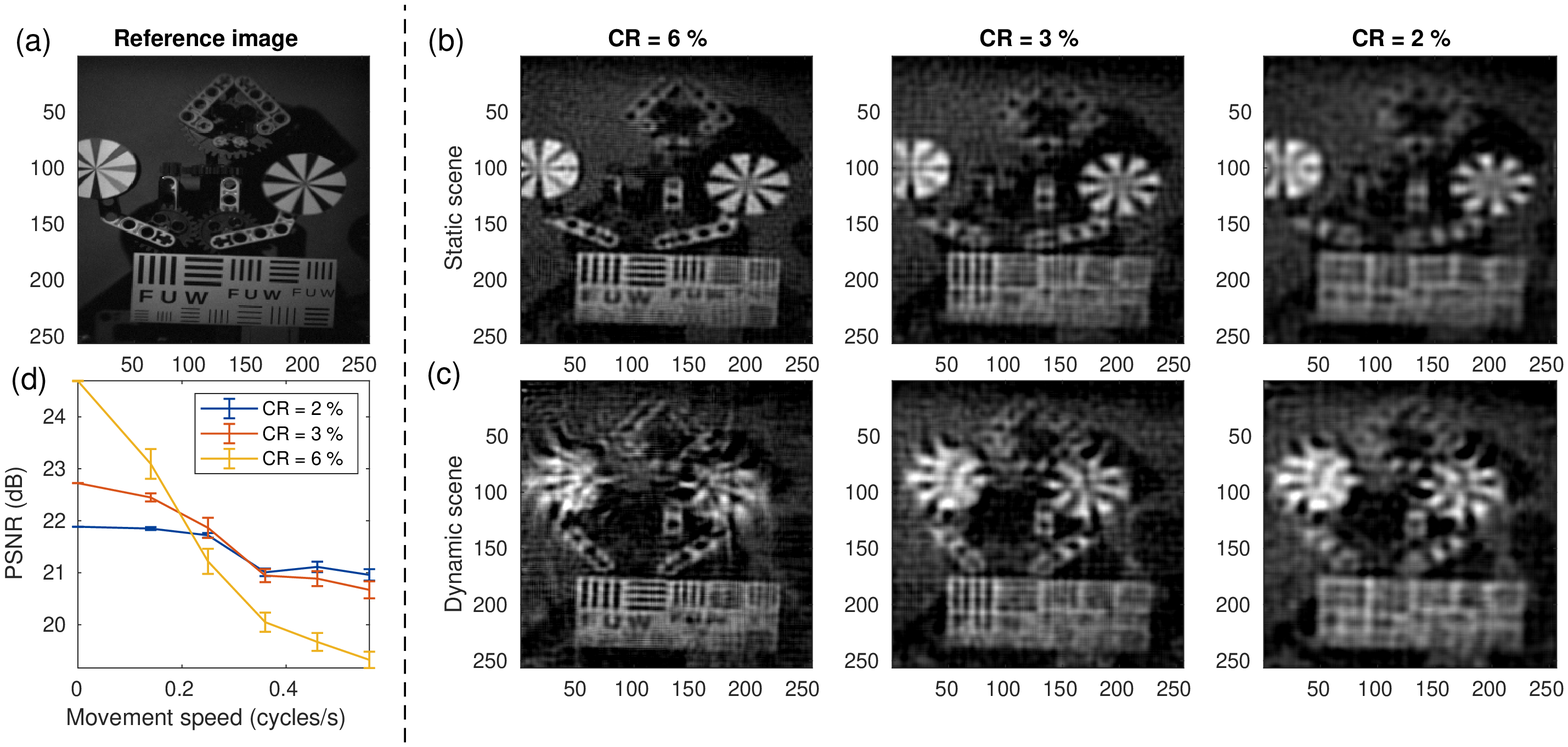}
	\caption{Examples of video frames obtained with the real-time single-channel visible SPI configuration using the proposed sampling with different compression ratios $CR=k/n$. (a)~Reference image measured with $CR = 100\%$ (robotic toy constructed with Lego bricks). (b)~Reconstructed images with $CR = 2, 3,$ or $6\%$ for still scene. (c)~Reconstructed video frames with $CR = 2, 3,$ or $6\%$ for moving scene. More detailed video reconstructions with varying movement speeds are presented in \textbf{Visualizations 4-6}. (d)~Average PSNR of the reconstructions obtained for each sampling ratio as a function of the movement speed of the scene. The reference image (a) is used as the ground truth for calculating PSNR, and the calculation is performed only for these frames, in which the position of the rotating wheels matches the one from the reference image.
	}
	\label{fig.lego}
\end{figure}

 The DMD (Vialux V-7001 XGA with DLP7000 chip) is operated at $22.7$~kHz and at a spatial resolution of $256\times256$ (with $3\times4$ pixels blocks assigned equal values). The infrared channel makes use of reflective optics. The detector (Vigo System, PVI-4TE-1x1mm) is a four-stage thermoelectrically cooled IR photovoltaic detector based on HgCdTe heterostructures optimised for the range 1-5.5$\mu$m. The detector is equipped with integrated transimpedance amplifier (DC 100k V/A). For the visible range, we use Thorlabs PDA100A2 amplified silicon photodiodes. The signals are digitized with a Picoscope 5000 DAQ with a sampling of $256$ns and streamed through a USB bus for real-time processing. A multiprocess Python program reads, synchronizes, and averages the data, and reconstructs images from two independent channels in real time. The flowchart in Fig.~\ref{fig.flowchart} shows the data processing stages in the two channels and image integration into a single pseudocolor frame.
 For polarization imaging, the two polarization channels provide the red and blue components of the red-green-blue color representation. For visible-infrared imaging, the visible channel is represented as a gray-scale image, and is replaced with the infrared channel represented with an orange image at those locations where the infrared channel exceeds a threshold value that corresponds to a temperature of approximately $100^{\circ}$C.
 
 Sample image reconstructions are shown in Fig.~\ref{fig.lutownica}. The following parameters were used $m=7,p=1,\mu=0.5, CR=6\%$. Respective Visualizations include sample movies with a birefringent ruler moved in front of a linear polarizer~(Visualization1 and Visualization2) as well as with a hot soldering iron tip moved in front of a resolution test image (Visualization3).
 
The compression ratios between $2\%$ to $6\%$ allow for imaging at in between $17$~Hz and $6$~Hz. For still objects, image quality naturally increases with the compression ratio $k/n$ which is inversely  proportional to the framerate. However, for non-stationary objects the value of PSNR is a trade-off between compression ratio and acquisition time. This trade-off is characterized in Fig.~\ref{fig.lego} showing a rotating Lego robot. Fig.~\ref{fig.lego}(a) and (b) show the SPI measurements of the robot at the compression ratio equal to $100\%$, $6\%$, $3\%$ and $2\%$, respectively. Rotating wheels give strong artifacts shown in Fig.~\ref{fig.lego}(c), and the PSNR drops the most rapidly with angular velocity when the compression ratio is the largest (See Fig~\ref{fig.lego}(d)). These results are also shown in Visualizations 4-6, made for a varying angular speed of robot wheels.

\begin{table}[t!]
\centering
\caption{Performance of D-FDRI reconstruction method in comparison with the compressive sensing (CS) techniques in terms of the average reconstruction time and PSNR obtained for compression ratios $CR  = 2\%, 3\%,$ or $6\%$. In both cases, the same measurement matrices have been used for each value of $CR$, and the measurements have been obtained using a single-channel visible SPI for the still robotic toy scene presented in Fig.~\ref{fig.lego}(a). CS reconstructions have been calculated via solving the basis pursuit denoise problem with the total variation regularisation using NESTA solver\cite{NESTA_2011}.	}
\begin{tabular}{l|ll|ll}
\hline
\multicolumn{1}{c|}{CR} & \multicolumn{2}{c|}{PSNR (dB)}                     & \multicolumn{2}{c}{\begin{tabular}[c]{@{}c@{}}Reconstruction\\ time (s)\end{tabular}} \\ \cline{2-5} 
\multicolumn{1}{c|}{}   & \multicolumn{1}{c}{D-FDRI} & \multicolumn{1}{c|}{CS} & \multicolumn{1}{c}{D-FDRI}                & \multicolumn{1}{c}{CS}                \\ \hline
2 \% & 21.95 & 21.86 & 0.019 & 6.51 \\
3 \% & 22.73 & 22.43 & 0.028 & 9.45 \\
6 \% & 24.76 & 24.45 & 0.056 & 21.7 \\ \hline
\end{tabular}
\label{table.fdri_vs_nesta}
\end{table}

Finally, in Table.~\ref{table.fdri_vs_nesta} we compare the proposed D-FDRI reconstruction method with the classical compressive sensing (CS) approach in terms of both image quality and reconstruction time. In both cases, the same differential binarized DCT measurement matrices are used with $m=7$ and $p=1$. For the purpose of CS reconstruction, we solve the basis pursuit denoise problem with the total variation regularisation using NESTA solver \cite{NESTA_2011}. As the differential measurement matrix $\bf{D}^p_k\cdot \bf{M}$ is not semi-orthogonal  (i.e. $(\bf{D}^p_k\cdot \bf{M})\cdot (\bf{D}^p_k\cdot \bf{M}) \neq \bf{I}$,
where $\bf{I}$ is an identity matrix), we reformulate the CS optimization problem into the following equivalent form:
\begin{equation}
\bf{\tilde{x}} \ = \ \underset{\bf{x}}{\textrm{argmin}} \ \lambda \,\|\bf{x}\|_{TV} + \tfrac{1}{2} \,\|  \bf{\Sigma}^{-1} \cdot \bf{U}^T \cdot \bf{D}^p_k \cdot \bf{y} - \bf{V}^T \cdot \bf{x}  \|_2^2,
\label{eq.TV_optim}
\end{equation}
where $\bf{D}^p_k\cdot \bf{M}=\bf{U} \cdot \bf{\Sigma} \cdot\bf{V}^T$ is the SVD decomposition of matrix $\bf{D}^p_k\cdot \bf{M}$. The parameter $\lambda$ in Eq.~(\ref{eq.TV_optim}) controls the trade-off between sparsity and fidelity of the reconstruction. Denotations 
$\|\bf{v} \|_{2}$ and $\|\ \bf{v} \|_{TV}$ stand for the $\ell_2$-norm and total-variation norm of a vector  $\bf{v}$, respectively.
The reconstruction times presented in Table.~\ref{table.fdri_vs_nesta} do not include either the time necessary to compute SVD for CS reconstruction or to compute matrix $\bf{P_g}$ for D-FDRI, as they both may be computed only once for each measurement matrix and stored. The reconstruction times have been obtained on an Intel i5-4590 CPU without any GPU acceleration.
For the given sampling, D-FDRI is faster than TV-norm minimization by over two orders of magnitude. As a matter of fact, being a single-step reconstruction method based on calculating a single matrix-vector product per image, D-FDRI is fast enough to reconstruct video images with resolution $256 \times 256$ in real-time on a desktop CPU using over 20~kHz DMD for image sampling. At the same time, both methods produce reconstructions of comparable quality, while D-FDRI yields on average $0.23$~dB higher PSNR owing to better robustness to noise. This result  does not prove that the proposed method would always give a higher PSNR than the methods of compressive sensing but it indicates that both approaches typically provide comparable results.

\section{Conclusions}
We propose a fast, differential, single-pixel imaging framework and demonstrate real-time SPI at the resolution of $256\times 256$ at frame-rates between $6$ and $17$~Hz in two independent channels assigned to different spectral bands or polarizations. The D-FDRI reconstruction method is based on a matrix-vector multiplication (one multiplication per channel per frame) and resembles the solution proposed in Ref.~\cite{Czajkowski_2018} but is modified to account for the differential treatment of the measured data with either a Laplace or gradient discrete operators.  It gives a similar or even better image quality as could be obtained with compressive sensing, but is significantly faster than iterative optimization techniques.
Measurement of the zeroth spatial frequency is decomposed into several measurements with binary patterns with help of an auxiliary  binary numerically optimized matrix. D-FDRI could use various bases of patterns, and we focus on binarized DCT basis with varied thresholding. The entropy of the detection signal is increased in comparison to standard SPI protocols enabling to reduce the bit-resolution requirements of the DAQ or otherwise improving noise robustness of the proposed SPI method. A trade-off behaviour between imaging quality and compression ratio is demonstrated as a function of the frame-rate.

\begin{backmatter}
\bmsection{Disclosures}
The authors declare no conflicts of interest. 

\bmsection{Data Availability Statement}
Source data will be provided by the authors at a reasonable request. Source code for calculating measurement and reconstruction matrices will be made public at~\cite{DFDRI} after article publication. The FDRI part of the code is based on~\cite{FDRI}. We have used the Caltech256 image database~\cite{CALTECH256} in this work.  

\bmsection{Funding}
National Science Center, Poland - (RS,PW,RK)-UMO-2017/27/B/ST7/00885, (AP)-UMO-2019/35/D/ST7/03781.
\end{backmatter}

\bibliography{biblio}

\end{document}